\documentclass[10pt,english,aps,notitlepage,twocolumn,shortbibliography,nofootinbib,prl,floatfix]{revtex4-2}
\usepackage{mathtools,amsthm,amsmath,amsfonts,amssymb}
\usepackage[colorlinks = true,
            linkcolor = blue,
            urlcolor  = blue,
            citecolor = blue,
            anchorcolor = blue, hypertexnames=true]{hyperref}
\usepackage{graphicx}
\usepackage{subfigure}
\usepackage{physics}
\usepackage{comment}
\usepackage{pgfplots}
    \usetikzlibrary{positioning}
\usepackage[normalem]{ulem}
\usepackage[noend]{algpseudocode}
\usepackage[dvipsnames]{xcolor}
\usepackage{orcidlink}
\usepackage{tikz}
    \usetikzlibrary{math}
\pgfplotsset{compat=1.18}

\DeclareMathOperator{\sJ}{\mathcal{J}}
\def\be{\begin{equation}}
\def\ee{\end{equation}}
\def\qq{\mathfrak{q}}
\def\e0{E_{0}}
\def\thetas{\theta_s}

\def\Sh{T}
\def\omegam{\omega_{0}}

\newcommand{\sect}[1]{\emph{#1}---}

\newcommand{\ictpsaifradd}{ICTP South American Institute for Fundamental Research \\
                           Instituto de F\'{i}sica Te\'{o}rica, UNESP - Univ. Estadual Paulista \\
                           Rua Dr. Bento Teobaldo Ferraz 271, 01140-070, S\~{a}o Paulo, SP, Brazil}

\begin{document}
\title{{Universality and Emergent Classicality in Krylov Space}}
\author{Debarghya Chakraborty\,\orcidlink{0009-0000-1078-5348}}
    \email{debarghya.chakraborty@ictp-saifr.org}
\affiliation{\ictpsaifradd}
\begin{abstract}
Krylov complexity grows linearly at late times for many systems. In this regime, we give a general spectral formula for the leading growth coefficients of Krylov complexity and its higher moments, without requiring detailed knowledge of the Lanczos coefficients. For states with a narrow energy distribution compared to the bandwidth, dynamics in the Krylov basis resembles motion along a classical trajectory. We demonstrate our results for the critical XY chain and double-scaled SYK. For the latter, we compute the leading late-time behavior of the chord-number moments through one-loop order.
\end{abstract}
\maketitle
\sect{Introduction}
Universality is a central organizing principle in modern physics. It refers to the phenomenon where a class of macroscopic systems with drastically different microscopic details obey the same scaling laws for coarse-grained observables. Renormalization-group fixed points provide some of its most striking examples. Another example is the level-spacing statistics of generic nonintegrable quantum systems, described by random-matrix ensembles. In this Letter, we present a notion of universality for late-time dynamics in Krylov space that applies to a broad class of physical systems. It describes the ballistic motion of a quantum state in a distinguished orthogonal basis. This motion is determined by the energy distribution of the state through a universal relation.
    
We call this basis \textit{the} Krylov basis following standard terminology. The Krylov basis relevant for describing time-evolution $\ket{\psi(t)} := e^{-i H t} \ket{\psi(0)}$ is constructed through the recursion
\be
H \ket{n} = b_{n-1} \ket{n-1} + a_{n} \ket{n} + b_{n} \ket{n+1}, \label{eq:orthonormal_Lanczos_H}
\ee 
with boundary conditions $b_{-1}=a_{-1} = 0$ and $\ket{0}:= \ket{\psi(0)}$, such that the basis states $\{ \ket{n} \}$ are orthonormal. The Hamiltonian $H$ acts like a tridiagonal matrix in this basis, i.e., $M_{mn} := \bra{m} H \ket{n} = b_{n-1} \delta_{m,n-1} + a_n \delta_{m,n} + b_{n} \delta_{m,n+1}$. The Lanczos coefficients $b_n$, $a_n$ are fixed by orthogonality, in terms of $\bra{\psi} H^{k} \ket{\psi}$ for $k$ up to $2n+2$. The Krylov basis is entirely determined by the survival amplitude 
\be C(t) := \bra{\psi(0)} \ket{\psi(t)}, \label{eq:Ct_def} \ee 
or equivalently, its Fourier transform 
\be \Phi(E) := \frac{1}{2\pi} \int^{\infty}_{-\infty} dt~C(t) e^{i E t} ,  \label{eq:Phi_E_def} \ee 
which we call the spectral function. In this basis the state is expanded as
\be \ket{\psi(t)} = \sum_{n=0} \varphi_n(t) \ket{n},\quad \varphi_n(t):=\bra{n}\ket{\psi(t)}, \label{eq:orthonormal_nbasis_expansion}\ee
where $\varphi_n(t)$ are determined in terms of $C(t)$. The moments of the distribution $\{\abs{\varphi_n(t)}^2\}$ are given by powers of the ``complexity'' operator $\hat{n} := \sum_n n \ket{n}\bra{n}$ in the state $\ket{\psi(t)}$:
\be \expval{\hat{n}^k(t)} = \sum_n n^k \abs{\varphi_n(t)}^2.  \ee 
The Krylov basis has emerged as a framework for characterizing quantum dynamics \cite{Dymarsky_2020, Caputa:2021sib, Dymarsky_2021, Balasubramanian_2022, Bhattacharya_2022, Bhattacharjee_2022, Balasubramanian:2023_plateau, Erdmenger_2023, Angelinos:2025drf, Takahashi_2025, grabarits_2026}. This activity was stimulated by \cite{Parker_2019} which argued that operator wavefunctions spread universally in the Krylov basis under Heisenberg evolution, and that $\expval{\hat n(t)}$ grows exponentially for local operators in generic nonintegrable systems. The mean $\expval{\hat{n}(t)}$, called Krylov complexity is a simple measure of operator growth obtained from the  operator autocorrelation function. In general, Krylov complexity shows rich phenomenology across several time scales \cite{rabinovici_review_complexity, Nandy_2025}. 

\sect{Statement}
For a broad class of systems where $\Phi(E)$ has bounded support, $\expval{\hat{n}(t)}$ is expected to exhibit a parametrically long regime of linear growth \cite{Barb_n_2019,Jian_2021, Rabinovici_2021,Kar_2022}. Extended linear growth is also a central feature of holographic proposals relating quantum complexity to the growth of Einstein-Rosen bridges \cite{susskind2014, Stanford_2014}. In finite systems, a maximum energy directly leads to $\Phi(E)$ vanishing beyond a cutoff. In sufficiently large systems, this regime may persist for times exponential in system size, before a slow drift sets in and eventually leads to saturation. For infinite systems with bounded spectral support, the linear growth may persist indefinitely.

We shift the energies by a constant so they lie in $[-\e0, \e0]$. We assume scenarios where $\Phi(E)$ is exactly, or after coarse-graining, smooth and continuous, vanishing for $\abs{E} > \abs{\e0}$ with no gaps inside $[-\e0,\e0]$. The generating function of $\expval{\hat{n}^k(t)}$ was studied in \cite{Chakraborty:2026ssu} which found that at late times
\be \expval{\hat{n}^k(t)} = v_k t^k + O(t^{k-1}), \label{eq:ballistic_scaling} \ee 
where the slopes $v_k$ have the spectral representation   
\be 
v_k = \int^{\e0}_{-\e0} dE~\Phi(E) \big( \e0^2 - E^2  \big)^{\frac{k}{2}}, \label{eq:vk_from_Phi}
\ee 
which can also be rewritten in the time-domain 
\be v_k = \frac{\e0^{k+1} \Gamma
   \left(\frac{k+2}{2}\right) }{\sqrt{\pi 2^{1-k}} } \int^{\infty}_{-\infty} dt~C(t) \frac{J_{\frac{k+1}{2}}(\e0 t)} { (\e0 t)^{\frac{k+1}{2}} }, \label{eq:vk_from_Ct} \ee 
if $\e0$ is known, but only $C(t)$ is directly accessible. 
Ref. \cite{Chakraborty:2026ssu} derived \eqref{eq:ballistic_scaling} using the propagation generated by $\hat{n}$ in the energy basis. The main technical ingredient in the derivation is the asymptotic form of $\ket{n}$ in the energy basis \eqref{eq:szego_asymptotic_pn}, that follows from the orthogonal polynomial asymptotics found by Szeg\H{o} \cite{szego1959orthogonal}. The $O(t^{k-1})$ bound on corrections requires further assumptions but is expected to be generic. The asymptotics \eqref{eq:szego_asymptotic_pn} also offer a way to directly approximate $\varphi_n(t)$ as explored in Sec.~\ref{sec:krylov_probabilities}. While completing this paper, we became aware of independent derivations of \eqref{eq:vk_from_Phi} in recent works \cite{alfinito2026higherloopwormholelengthsinedilaton,qu_2026_geodesic}, for general $k$ and for $k=1$, respectively.

A simpler physical argument is the following. In our conventions, $\lim_{n \rightarrow \infty} b_n = \e0/2$ and $\lim_{n \rightarrow \infty} a_n = 0$. The Lanczos coefficients approach their asymptotic value at a scale $n_{c}$. If this approach is algebraic $\sim n^{-\gamma}$ for $\gamma>0$, $n_c(\epsilon)$ can be defined as $\abs{b_{n_c(\epsilon)} - \frac{\e0}{2}} = \epsilon$. Sometimes $n_c$ can be estimated as the $n$ at which the saddle point of $\int dE\,E^{2n}\Phi(E)$ reaches $\abs{E}=\e0$, which depends on the detailed profile of $\Phi(E)$. 

The equations of motion for the Heisenberg picture $\hat{n}(t)$ are 
\be \frac{d \hat{n}(t)}{dt} = e^{i M t} B e^{-i M t}, \quad \frac{d^2 \hat{n}(t)}{dt^2} = e^{i M t} \mathcal{F} e^{-i M t}, \label{eq:n_operator_eqs_of_motion}  \ee 
expressed in terms of $B:=i[M,\hat{n}]$, $\mathcal{F} :=-[M,[M,\hat{n}]]$. We denote the asymptotic form of operators at indices $n\gg n_c$ using the subscript $_{n_c}$. $M_{n_c}$ is a constant symmetric tridiagonal matrix with off-diagonal entries $\e0/2$ and zero along the diagonal. $B_{n_c}$ is $i$ times the antisymmetrized $M_{n_c}$. This implies
\be 
\mathcal{F}_{n_c}\simeq 0,\qquad
B_{n_c}^2\simeq \e0^2-M_{n_c}^2, \label{eq:bulk_asymptotic_operator}\ee 
where in writing \eqref{eq:bulk_asymptotic_operator} we neglect boundary terms and deviations of the Lanczos coefficients from their asymptotic values. Since the wavepacket moves towards larger $n$, we take the positive root 
\be 
B_{n_c}\simeq \sqrt{\e0^2-M_{ n_c}^2}. \label{eq:asymptotic_B} \ee 
Once the wavepacket $\varphi_n(t)$ has propagated into this region $n \gg n_c$, \eqref{eq:bulk_asymptotic_operator}, \eqref{eq:asymptotic_B} can be used to approximate the solution of \eqref{eq:n_operator_eqs_of_motion} within expectation values. Vanishing of $\mathcal{F}_{n_c}$ implies the ballistic growth in \eqref{eq:ballistic_scaling}. Using energy conservation of $\ket{\psi(t)}$, the asymptotic relation \eqref{eq:asymptotic_B} gives rise to the spectral relation \eqref{eq:vk_from_Phi}. 

The onset time $t_{\text{on}}^{(k)}$ for the validity of \eqref{eq:ballistic_scaling} depends on $k$, but this effect is small \cite{Chakraborty:2026ssu}, so we restrict to the single scale $t_{\text{on}}$. Subleading corrections in \eqref{eq:ballistic_scaling} encode details of the transient propagation and the approach of the Lanczos coefficients to their asymptotic values. The width of $\Phi(E)$ sets the scale for Lanczos coefficients; if this scale does not depend on the system size but $\e0$ does then $t_{\text{on}}$ grows with the system size.

\sect{Emergent classicality}
When \eqref{eq:ballistic_scaling} is valid for sufficiently large times $\expval{\hat{n}^k(t)}/(v_k t^k)$ approaches $1$ and the Krylov distribution is simply and universally characterized by the spectral representation of $\ket{\psi}$ through \eqref{eq:vk_from_Phi}. In this regime, there is a simple relation between the width of $\ket{\psi(t)}$ in the Krylov and energy bases. In many contexts, the width of $\Phi(E)$ is subextensive but $\e0$ is extensive. If $\Phi(E)$ is sharply peaked about its mean $\mu_1$, expanding $(\e0^2 -E^2)^{k/2}$ about $\mu_1$ gives $v_{k} \sim (\e0^2 - \mu^2_1)^\frac{k}{2}$ with corrections that are suppressed by the system size. 

$\ket{\psi(t)}$ becomes effectively classical in the Krylov basis in the following sense. $\expval{\hat{n}^{k}(t)} \approx \expval{\hat{n}(t)}^k$, which means that $\hat{n}$ becomes a classical macroscopic variable. Smooth functions of $\hat{n}$ become deterministic, evaluated at late times. The Krylov probabilities move along a ballistic trajectory determined by \eqref{eq:ballistic_scaling}. 

\begin{figure}[ht]
  \centering
  \includegraphics[width=0.44\textwidth]{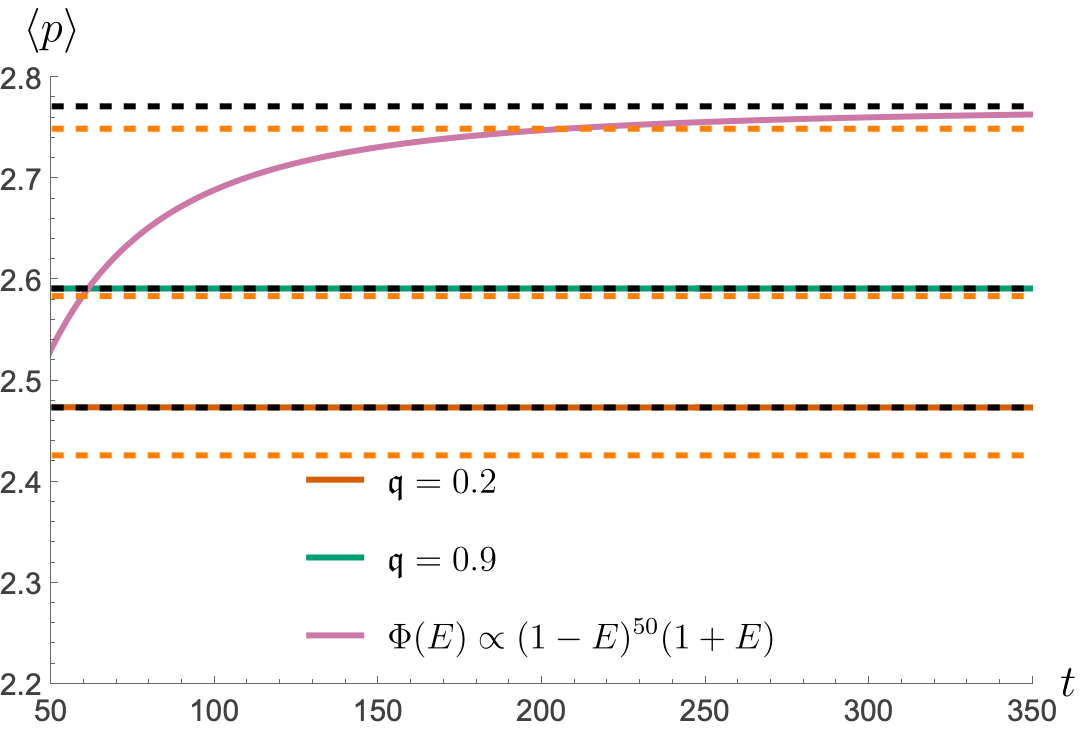}
   \caption{Saturation of $p$: for DSSYK complexified time evolution \eqref{eq:complexified_k_complexity} at $\beta=4, \sJ=1$ (and $\qq$ as indicated) and evolution generated using toy measure $\Phi$ (magenta). Black dashed lines are evaluations of \eqref{eq:mean_phase}. The orange dashed lines are $\arccos(\mu_1/\e0)$, which serves as a good approximation for the narrow distribution at $\qq=0.9$. $\expval{p}$ saturates later for the toy measure because of larger $t_{\text{on}}$. }
  \label{fig:phase}
\end{figure}

Surprisingly, the phases in $\varphi_n(t)$ \textit{also} become deterministic. Despite the obstruction to defining a global phase operator conjugate to $\hat{n}$, a simple workaround is to use the shift operator $\Sh = \sum_n \ket{n+1}\bra{n}$ \cite{susskind_glogower}. The identification $\Sh = e^{i p}$ is valid when $\ket{\psi(t)}$ has negligible support close to $n=0$. $\arg \expval{\Sh(t)}$  can be interpreted as the mean $\expval{p}$.
Because $M_{n_c} + i B_{n_c} \approx \e0 \Sh$, the $p$ distribution is approximately $t$-independent and 
\be \expval{p} \approx \arg \frac{1}{\e0} \int dE \Phi(E) \big( E + i  \sqrt{\e0^2 - E^2} \big), \label{eq:mean_phase} \ee 
with analogous diagnostics for fluctuations of $p$. When $\Phi(E)$ is sharply peaked, so is the $p$ distribution and $\expval{p}$ can be approximated by  $\arccos(\mu_1/ \e0)$, see Fig.~\ref{fig:phase}. This, together with $\abs{\expval{\Sh(t)}} \approx 1$ implies an approximately linear winding phase for $\varphi_n(t)$ and can be seen as one possible mechanism underlying Krylov winding studied in \cite{Perugu_2026}.  

Thus, at late times $\ket{\psi(t)}$ describes a well-defined trajectory $\big( \sqrt{\e0^2 - {\mu^2_1}} t~,~ \arccos(\mu_1/ \e0) \big)$ for the  semiclassical variables $n$ and $p$. This is similar to a coherent state with large mean occupation number and a well-defined phase. The number-phase uncertainty is respected because variance of $\hat{n}$ is not small in absolute terms.      
\begin{figure*}[ht]
  \centering
  \includegraphics[width=0.37\textwidth]{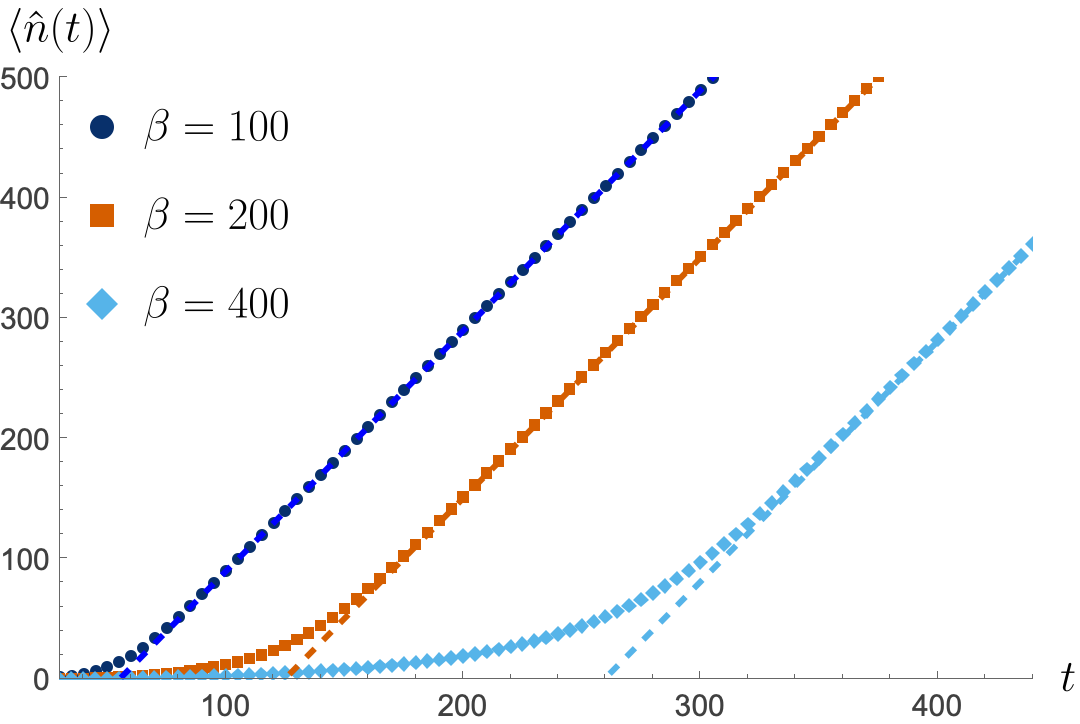}\hfill
  \includegraphics[width=0.40\textwidth]{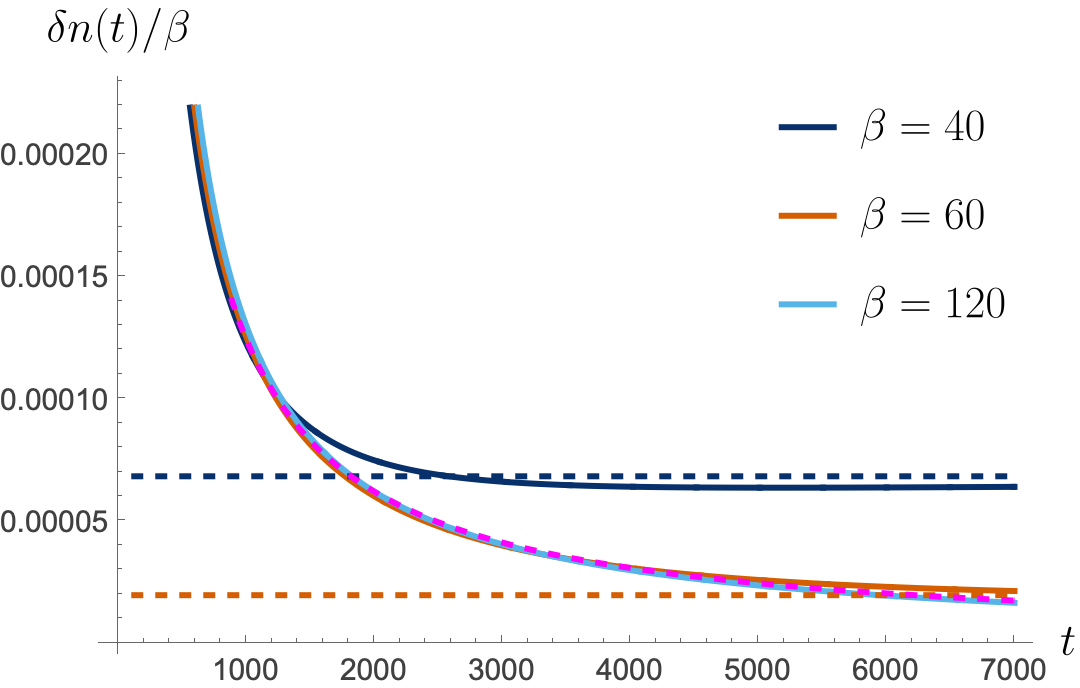}
   \caption{Critical XY model. (a) Onset: Dashed lines show linear fits (consistent with \eqref{eq:ansatz_for_intercept_exp}) where the slopes are computed from \eqref{eq:vk_from_Ct}. (b) Saturation of $\delta n(t)$: Horizontal dashed lines are $\sqrt{v_2-v^2_1}/(\beta v_1)$ paired with corresponding $\beta$. Multiplication by $1/\beta \propto \alpha$ leads to a collapse of $\delta n(t)$ along dashed magenta curve decaying as $1/t$.   }
  \label{fig:xy_model}
\end{figure*}
In fact, the variance of $\hat{n}$ \textit{grows} as $t^2$ but it is the relative width    
\be \delta n(t) = \frac{\sqrt{\expval{\hat{n}^2(t)} - \expval{\hat{n}(t)}^2}}{\expval{\hat{n}(t)}}, \ee
that converges to $ \big(\sqrt{v_{2} - v^2_{1}} \big)/v_1$. The approach of $\delta n(t)$ to its $t \rightarrow\infty$ asymptotic defines the scale $t_{\text{cl}}$, marking the point where the subleading terms in \eqref{eq:ballistic_scaling} become negligible for $k=1,2$.     

There are physical scenarios where the sharpness of Krylov distribution does \textit{not} hold. Take $\ket{\psi}$ uniformly populated within an energy window of width $2 \e0$. $\e0$ could be taken finite to mimic a microcanonical state in a large system. Then \eqref{eq:ballistic_scaling} holds for exponentially long times $t$ with $O(1)$ values of $v_k$ and $\delta n(t)$ asymptotes to $\approx 0.284$. Similarly  $\lim_{t\rightarrow\infty} \delta n(t) \approx 0.202$ for the dynamics of a fully delocalized state under a Gaussian random matrix of large size assuming that coarse-grained $\Phi(E)$ converges to the Wigner semicircle distribution. The narrowness condition is also not sufficient if $\Phi(E)$ is sharply peaked very close to $\e0$.

Conversely, a narrow $\Phi(E)$ away from the edges $\pm \e0$ is sufficient but not necessary for the wavepacket to become sharply peaked in the Krylov basis. $\delta n(t)$ decays as $\frac{1}{\sigma t}$ when $\Phi(E) \propto e^{-\frac{E^2}{2 \sigma^2}}$, while $\expval{\hat{n}(t)} = \sigma^2 t^2$. 

\sect{Operator growth}
Here we estimate $t_{\text{on}}$ and $t_{\text{cl}}$ using a physically motivated model for $\Phi(\omega)$ relevant for operator growth. Following \cite{Parker_2019}, we assume that $\Phi(\omega)=\Phi(-\omega)$ decays as $\sim e^{-\frac{\pi}{2 \alpha} \abs{\omega}}$ in the bulk, and that it vanishes at an extensive maximal frequency $\omegam$. Accordingly, our analysis relies exclusively on $\omegam$ and $\alpha$. We assume that $\Phi(\omega)$ can be treated as a smooth function and ignore additional structure such as its precise behavior near $\omegam$. For symmetrically ordered $C(t)$ at finite inverse temperature $\beta$, it is expected that $\alpha \sim \frac{\pi}{\beta}$. To leading order in  $\alpha/\omegam \ll 1$, $v_2 - v_1^2 \sim \alpha^4/\omegam^2$ and $v_1 \sim \omegam$.

Matching $b_n \approx \alpha n$ for $n \lesssim n_c$ and $b_n = \omegam/2$ for $n\geq n_c$, we get $n_{c} \approx \omegam/2 \alpha$. Using the known transient exponential growth $e^{2 \alpha t_{\text{on}}} \approx n_c$ we get
\be t_{\text{on}} \approx \frac{1}{2 \alpha} \log(\frac{\omegam}{\alpha}).  \label{eq:estimate_t_on} \ee
To find $t_{\text{cl}}$, we need control over the subleading terms in \eqref{eq:ballistic_scaling}. In the End Matter we describe a phenomenological model designed with input from numerical experiments that gives
\be \delta n(t) \sim  \frac{1}{\alpha t}, \quad t_{\text{on}} \ll t \ll t_{\text{cl}} \label{eq:1overt_decay}. \ee
Matching \eqref{eq:1overt_decay} with its asymptotic limit $\sim \alpha^2/\omegam^2$ gives $t_{\text{cl}} \sim \omegam^2/\alpha^3$.  

As a concrete physical example, we study the two-point function \eqref{eq:xy_t_domain} of the transverse magnetic field in the critical spin-$1/2$ XY model (also studied in \cite{Avdoshkin_2024,Angelinos:2025drf}). This is a \textit{free} system with $\Phi(\omega) \sim e^{-\beta \abs{\omega}/2}$ and fixed $\omegam = 2$. The saturation of $\delta n(t)$ becomes numerically accessible because $\omegam$ is a fixed $O(1)$ number. It mimics features of generic systems as $\omegam \gg \alpha = \frac{\pi}\beta$ up to an overall rescaling of time. Fig.~\ref{fig:xy_model} illustrates $t_{\text{on}}$ and $t_{\text{cl}}$ consistent with our model in this system.  \begin{figure*}[ht]
  \centering
  \includegraphics[width=0.36\textwidth]{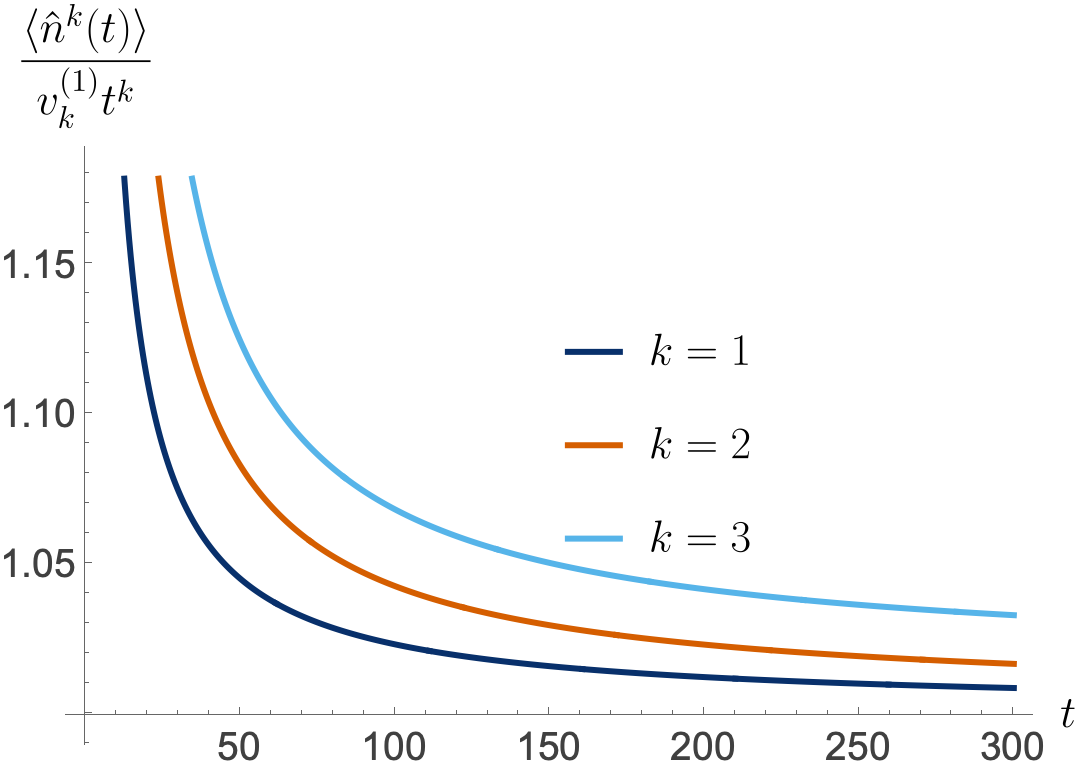}\hfill
  \includegraphics[width=0.36\textwidth]{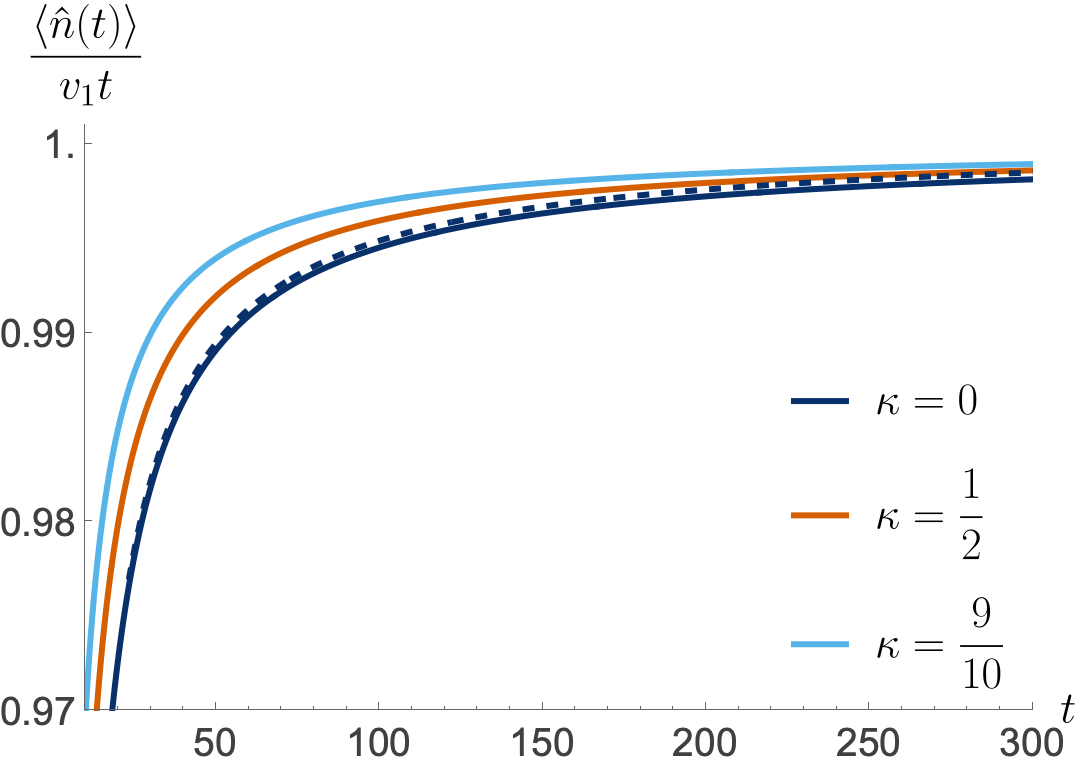}
  \caption{(a) DSSYK Complexity at $\beta=10$: Ratio of $\expval{\hat{n}^k(t)}$ to $v^{(1)}_k t^k$ using analytic one-loop $v^{(1)}_{k}$ computed in \eqref{eq:1_loop_corr_vk}. (b) Deformed DSSYK: Ratio of $\expval{\hat{n}(t)}$ to $v_1 t$ for a one-parameter deformation labeled by $\kappa$ \eqref{eq:one_param_def} at $\beta=2$. $\kappa=0$ corresponds to undeformed DSSYK. The dashed line uses the one-loop $v^{(1)}_k$ and solid lines are numerically integrated $v_k$. $\qq=0.9$ and $\sJ=1$ in both panels.}
  \label{fig:dssyk_vk}
\end{figure*}

\sect{Double-Scaled SYK (DSSYK)}
\label{sec:dssyk}
The SYK model \cite{Sachdev_Ye,Maldacena_2016} of $N$ Majoranas with random $q$-body interactions has a solvable double-scaling limit $N \rightarrow \infty$ and $q \rightarrow \infty$ keeping $\lambda := \frac{2q^2}{N} =:-\log \qq $ fixed \cite{Berkooz_2018,Berkooz_2019, Berkooz_2025_cordial}. At low energies, SYK is dual to JT gravity \cite{Maldacena_2016}, an effective theory of quantum gravity in two dimensions. Sine-dilaton gravity extends this description to DSSYK beyond the low-energy regime \cite{Blommaert_2025,blommaert2025_entropic}.

The Krylov basis constructed from the infinite-temperature thermofield double, using disorder-averaged moments, coincides with the chord-number basis \cite{Lin_2022, Rabinovici:2023yex, Ambrosini:2024sre}. The chord number gives a notion of a discrete length in the bulk and helps to understand the emergence of a bulk Hilbert space, with semiclassical features when $\lambda \rightarrow 0$. The dictionary between bulk length and chord number is given by \cite{Lin_2022, Rabinovici:2023yex, Xu:2024gfm, Ambrosini:2024sre, Heller_2025, Aguilar-Gutierrez:2025pqp}
\be  \expval{\tilde{l}(t)} = \lambda \expval{\hat{n}(t)}_{\beta}. \ee
$\expval{\hat{n}(t)}_{\beta}$ is the Krylov complexity of complexified time evolution, incorporating the Euclidean segment required to prepare the thermofield double. To characterize the corresponding normalized Krylov distribution, we consider the moments
\be \expval{\hat{n}^{k}(t)}_{\beta} := \frac{1}{Z(\beta)}  \bra{\psi(0)} e^{-(\frac{\beta}{2} - i t)H } \hat{n}^k e^{- (\frac{\beta}{2}  + i t) H }\ket{\psi(0)}. \label{eq:complexified_k_complexity} \ee  
The $\Phi(E)$ relevant for building the Krylov basis is the SYK density of states. In the spectral representation, the Euclidean evolution amounts to replacing $\Phi(E)$ by the normalized thermal measure $Z(\beta)^{-1}e^{-\beta E}\Phi(E)$, while keeping the Krylov basis fixed, see \cite{bhattacharya2026quadratic} for a discussion. 

$\expval{\tilde{l}(t)}$ can be interpreted as the expectation value of the geodesic length of a wormhole in an effective two-sided $\text{AdS}_2$ black hole geometry whose inverse temperature $\beta_{\text{BH}}$ differs from the DSSYK inverse temperature $\beta$, with the two related as \cite{Blommaert_2025}
\be \beta = \frac{2 \thetas- \pi }{\sJ \sin(\thetas)}, \quad  \beta_{\text{BH}}=\frac{\pi}{\sJ \sin(\theta_s)}. \label{eq:dssyk_saddle} \ee
$\sJ>0$ is an overall energy scale present in the SYK Hamiltonian. $\theta_s$ parameterizes the semiclassical saddle. From the Krylov perspective, $\theta_s$ is the asymptotic value of $\expval{p(t)}$. In the semiclassical limit,
\be \expval{\tilde{l}(t)} = 2 \log(\frac{ \sJ \beta_{\text{BH}}}{\pi} \cosh(\frac{\pi t}{\beta_{\text{BH}}} )), \label{eq:semiclassical_length} \ee 
with length fluctuations suppressed because $\expval{\hat{n}^{k}(t)}_{\beta} = \expval{\hat{n}(t)}^{k}_{\beta}$ to leading semiclassical order. A bulk prescription for $\expval{\hat{n}^{k}(t)}_{\beta}$ was proposed in \cite{Fu_2026}, while their quantum fluctuations at infinite temperature were studied through higher-loop expansions in \cite{alfinito2026higherloopwormholelengthsinedilaton}.

\eqref{eq:vk_from_Phi} provides a direct route to the late-time growth of all chord-number moments, including their quantum corrections. The loop expansion in $v_k$ directly follows from the loop expansion in $\lambda$ of the thermal energy moments, see \eqref{eq:vk_central_moment_to_loop} and the following discussion. The leading answer is
\be \bigg( \frac{\lambda}{t}  \bigg)^k \expval{\hat{n}^{k}(t)}_{\beta} \approx (2 \sJ \sin(\theta_s ) )^k =  \bigg( \frac{2 \pi}{\beta_{\text{BH}}} \bigg)^k,  \ee
agreeing with \eqref{eq:semiclassical_length} for $t \gg \beta_{\text{BH}}$. In this example $t_{\text{on}} \sim O(\lambda^0)$ due to extensive energy variance \cite{Rabinovici:2023yex,Chakraborty:2026ssu}.  
The one-loop answer $v^{(1)}_{k}$ is given in \eqref{eq:1_loop_corr_vk}. For $k=1$, it reproduces the large-$t$ value of one-loop $\expval{\tilde{l}(t)}/t$ computed in \cite{Okuyama_2023, Goel:2023svz} after analytic continuation. We show the numerical agreement with \eqref{eq:ballistic_scaling} in Fig.~\ref{fig:dssyk_vk} where we also include a one-parameter deformation widely studied in recent times in the context of end-of-the-world branes \cite{Okuyama:2023byh,Aguilar-Gutierrez:2026nmd,rajgadia2026}. 

\sect{Discussion}
The heuristics presented in this letter have the merit that they plausibly apply to large, finite systems. There is a parametrically long regime in $n$ where the Lanczos coefficients are expected to be well-described by fluctuations about smooth functions $b(x)$ and $a(x)$ sampled at $x=n/D$, where $D$ is the Krylov space dimension, exponential in system size. The constant Lanczos sequence we utilized emerges from the Taylor expansion $b(\frac{n_c}{D}) + (x -\frac{n_c}{D})\partial_x b + \ldots$. The gradient in the Lanczos coefficients remains parametrically small for $n\ll D$, suggesting an exponentially long window with growth $\expval{\hat{n}^k(t)} \sim v_k t^k$. 
The size and correlations of the fluctuations play a crucial role. For random matrix ensembles, the former decays as $1/\sqrt{D}$ \cite{Dumitriu:2002ntg, balasubramanian2022_bnfromrho}, which is favorable for the scaling \eqref{eq:ballistic_scaling} to persist for exponentially long times. In real systems, the fluctuations can be much larger and are known to impede propagation in the Krylov basis---like for operator dynamics in integrable systems \cite{Rabinovici_2022}. Since the average $b_n$ is extensive, there could be much larger fluctuations and yet our results could still hold over a time window polynomial in the system size. It is an interesting problem to quantify the size of fluctuations that would be compatible with an approximate form of \eqref{eq:vk_from_Phi}. 

The more intriguing point is the possible robustness of the relation between the width of $\Phi(E)$ and the sharpness of the Krylov distribution. It could very well extend far beyond cases where either the $\expval{\hat{n}^k(t)} \sim v_k t^k$ scaling or its predicted coefficients \eqref{eq:vk_from_Phi} are reliable. A preliminary numerical study is presented in Sec.~\ref{sec:finite_size_numerics}.

In the semiclassical limit of DSSYK, $p$ along with $\lambda \hat{n}$ become phase-space variables of the dual theory. Our work suggests that a similar classical picture can emerge dynamically at late times in more general systems. In this regime $\hat{n}$ and $p$ become concentrated around a single trajectory if the state has a narrow spectral spread. This raises the question of \textit{when} such an emergent classical description is physically meaningful. Our perspective complements recent studies of classicality in Krylov space using different approaches \cite{semiclassicalRMT,balasubramanian_wignernegativity, 2025_krylovwignerfunction, basu2025_wignernegativity,Murugan:2026yyu}. 

In DSSYK the chord number operator is proportional to the microscopic size operator \cite{Lin_2022} within a time regime that overlaps with the regime studied in this work. In this window, $\expval{\hat{n}^k(t)}_{\beta}$ describes moments of the size distribution of the complexified time evolution operator $e^{-(\beta/2+it)H}$ \cite{Qi_2019}. Thus \eqref{eq:vk_from_Phi} determines the leading growth of its size moments, including their fluctuations, directly from the ensemble-averaged density of states. More generally, the operator Krylov complexity moments whose late-time behavior we studied are known to upper bound operator size moments \cite{Parker_2019}. Both examples we studied suggest a deeper connection between operator size and spectral data that involves the spectral edge $\e0$. Understanding the physical mechanism behind this connection and its broader applicability is an interesting problem for the future.

\sect{Acknowledgements} I thank B. Bhattacharjee, A. Dymarsky, S. Aguilar-Gutierrez, O. Parrikar and D. Rosa for  discussions that helped this work. I am supported by the S\~{a}o Paulo Research Foundation (FAPESP) through the grant 2024/13100-8. I acknowledge the use of ChatGPT and Claude for language feedback and LaTeX troubleshooting.
\bibliography{kcomplexity_scaling}
\clearpage
\newpage
\section*{End Matter}
\sect{Orthogonal polynomials and Szego Asymptotics}
 The Krylov vectors $\ket{n} = p_n(H) \ket{0}$ where $p_n(x)$ is the orthonormal polynomial of order $n$ satisfying the same recursion as \eqref{eq:orthonormal_Lanczos_H} with $p_0(x)=1$. If the integral 
 \be  \int^{\e0}_{-\e0} dE \frac{\log \Phi(E) }{\sqrt{\e0^2- E^2}}, \ee 
 exists, the $n \gg 1$ asymptotics are given by
\be p_n(\e0 \cos(\theta)) = \sqrt{\frac{2}{\pi \Phi(\e0 \cos(\theta)) \e0 \sin(\theta) }} \cos(n \theta + \gamma(\theta) ),  \label{eq:szego_asymptotic_pn}\ee 
in terms of $\cos(\theta) = E_0/E$.
$\gamma(\theta)$ is also calculable in terms of $\Phi(E)$ but we ignore its precise value in this work.  \\~\ 

The propagator associated with the amplitude $\bra{E_1} e^{i \hat{n} \phi} \ket{E_2}$ underlies the physical time-evolution of the global mode in the bipartite scar construction of \cite{Chakraborty:2025bhk}.
\cite{Chakraborty:2026ssu} derived \eqref{eq:ballistic_scaling}, \eqref{eq:vk_from_Phi} by computing an approximation of this propagator using \eqref{eq:szego_asymptotic_pn}. The integral resulting from inserting this approximate propagator in the generating function $\bra{\psi(t)} e^{i \hat{n} \phi} \ket{\psi(t)}$ can be evaluated in the $t \rightarrow \infty$ limit using the stationary phase method. Expanding in small $\phi$ isolates the $v_k t^{k}$ piece and gives the explicit form  \eqref{eq:vk_from_Phi}.
\subsection*{DSSYK Complexity}
\label{sec:end_matter_dssyk}
After ensemble averaging, the energy spectrum of DSSYK is given by
\be E(\theta)= \frac{2\sJ \cos(\theta)}{\sqrt{\lambda (1-\qq)}},\quad \theta \in [0, \pi],  \ee 
such that $\e0 =\frac{2\sJ}{\sqrt{\lambda (1-\qq)}}$. The density of states is also conveniently expressed in terms of $\theta$ as
\be \tilde{\Phi}(\theta) d\theta = \frac{1}{2 \pi} ( e^{-i 2 \theta}, \qq)_{\infty} ( e^{i 2 \theta}, \qq)_{\infty} ( \qq, \qq)_{\infty}, \label{eq:dssyk_Phi_E}\ee 
where $\tilde{\Phi}(\theta)=\e0 \Phi( \e0 \cos(\theta) ) \sin(\theta)$. $(a, \qq)_n$ stands for the $\qq$-Pochammer symbol $\prod^{n}_{j=1}(1-a \qq^{j-1} )$. The corresponding Lanczos coefficients are given by 
\be b_n = \sJ \frac{ \sqrt{1-\qq^{n+1}}}{\sqrt{\lambda (1-\qq)}},\quad a_n =0. \label{eq:dssyk_lanczos}\ee
There is a simple generalization at the level of an effective chord Hamiltonian description given by the measure 
\be \tilde{\Phi}(\theta) d\theta = \frac{1}{2 \pi} \frac{( e^{-i 2 \theta}, \qq)_{\infty} ( e^{i 2 \theta}, \qq)_{\infty} ( \qq, \qq)_{\infty} }{( \kappa e^{i \theta}, \qq)_{\infty} (\kappa e^{-i \theta}, \qq)_{\infty} }d\theta, \ee 
and Lanczos coefficients
\be b_n = \sJ \frac{ \sqrt{1-\qq^{n+1}}}{\sqrt{\lambda (1-\qq)}},~~a_n = \frac{\kappa \sJ}{\sqrt{\lambda  (1-\qq)}} \qq^n, \label{eq:one_param_def}  \\~\ \ee
which reduces to \eqref{eq:dssyk_Phi_E}, \eqref{eq:dssyk_lanczos} when $\kappa=0$. We restrict to $\abs{\kappa}<1$ to avoid bound-state contributions. 

Combining \eqref{eq:complexified_k_complexity} with \eqref{eq:vk_from_Phi} gives
\be v_k =  \frac{\e0^k}{Z(\beta)} \int  \tilde{\Phi}(\theta) e^{-\beta \e0 \cos(\theta) } \sin(\theta)^{k} d\theta. \label{eq:dssyk_vk_class} \ee 
\subsection*{Semiclassical Limit}
The semiclassical $\lambda \rightarrow 0$ limit when $\kappa=0$ is given by 
\be Z(\beta) \approx \sqrt{\frac{1}{2 \pi \lambda}}\int d\theta \exp( -\frac{1}{\lambda} F+ h), \ee
where following \cite{Okuyama_2023} we call
\be  F = 2 \bigg(\theta- \frac{\pi}{2} \bigg)^2 + 2 \sJ \beta \cos(\theta),\ee
\be h = \log(2 \sin(\theta) ) - \frac{\sJ \beta \cos(\theta)}{2}. \ee 
The saddle point is defined by $\partial_{\theta}F \big|_{\theta=\theta_s}=0$ giving rise to the first equation in \eqref{eq:dssyk_saddle}. It is conveniently parametrized using
\be u = \frac{\pi}{2} - \thetas, \quad \beta(u) = \frac{-2 u }{\sJ \cos(u)},\ee 
\eqref{eq:dssyk_saddle} and \eqref{eq:dssyk_vk_class} imply the classical result $v_k = \big(\frac{2 \sJ}{\lambda} \sin(\thetas) \big)^k$. 

The one-loop $Z(\beta)$ is given by \cite{Goel:2023svz, Okuyama_2023}
\be \frac{\cos(u)}{\sqrt{1 + u \tan u}}\exp(- \frac{2 u}{\lambda} \big( u - 2 \tan u \big) + u \tan u ). \label{eq:one_loop_zbeta}  \ee
The one-loop answer is accurate away from the edges, i.e., $\lambda \ll \pi - \thetas$. Using \eqref{eq:one_loop_zbeta} we write 
\begin{subequations}
  \begin{align}
   \mu_{1} &= \frac{2 \sJ \sin u}{\lambda} +  \mu^{(1)}_{1} + O(\lambda), & \\
    \mu^{(1)}_1  &= \frac{\sJ  \left(2 u^2 \tan (u)+u-\sin (u) \cos
   (u)\right)}{4 \cos(u) (u \tan (u)+1)^2},  \label{eq:e_1loop}  \\
    \mu_{2} - \mu^2_{1} &= \frac{\sJ^2 \cos ^2(u)}{\lambda  (u \tan(u)+1)} + O(1), \label{eq:evar_1loop}&
      \end{align}
\end{subequations}
where $\mu^{(1)}_{1}$ is the $O(1)$ contribution to the energy. To compute the $O(\lambda)$ correction to $\lambda^k v_k$, it is sufficient to keep $\mu_1$ up to $\mu^{(1)}_{1}$ and energy variance to $O(\frac{1}{\lambda})$.

This can be seen by expanding $z_k(E) = (\e0^2 - E^2)^\frac{k}{2}$ around $\mu_1$  
\be \lambda^k v_k = \frac{\lambda^k}{Z(\beta)} \sum_r z^{(r)}_{k}(\mu_1)  \int \Phi(E)  (E-\mu_1)^r e^{-\beta E} dE, \label{eq:vk_central_moment_to_loop}  \ee
where $z^{(r)}_{k}$ is the $r$th derivative of $z_k$. It is easy to verify that $z^{(r)}_{k}(E)/z_{k}(E)$ scales as $\lambda^{r}$ by switching variables to $\epsilon \rightarrow \lambda E$ and by restricting to $\thetas$ away from the edges. From the moment-cumulant formula, it can be seen that the central moment in \eqref{eq:vk_central_moment_to_loop} comes with a leading factor of $\lambda^{-\lfloor r/2 \rfloor}$, which means that $r$th term in \eqref{eq:vk_central_moment_to_loop} contributes at $\lambda^{\lceil r/2 \rceil}$, taking $r \geq 2$. To obtain \(\lambda^k v_k\) through \(O(\lambda^p)\), it suffices to truncate the expansion on the rhs of \eqref{eq:vk_central_moment_to_loop} at $r=2p$, while evaluating the semiclassical free energy \(\log Z(\beta)\) up to \(O(\lambda^{p-1})\). The one-loop result computed using \eqref{eq:one_loop_zbeta} is 
\be \begin{split}
&v^{(1)}_{k} = \big( \frac{2 \sJ \cos(u) }{\lambda}  \big)^k  \bigg(1 +  \lambda k  \quad \\ 
 &\frac{  
\sin ^2(u) ((k-1) u \tan (u)+k)+2 u
   \tan (u)+1}{8 \cos^2(u)
   (u \tan (u)+1)^2} \bigg).
\end{split} \label{eq:1_loop_corr_vk}
\ee 

\subsection*{Operator growth \label{sec:phi_for_operator_growth}}
\sect{Estimating the timescales}
The heuristic derivation in the main text suggests the approximate \textit{operator} equation 
\be \hat{n}(t) \approx B_{n_c} t + X, \label{eq:operator_ansatz} \ee 
valid for the physical trajectory in the relevant regime, where crucially $X$ is time-independent. Matching the solution at $t=t_{\text{on}}$ gives 
\be  \expval{\hat{n}(t)} = v_1 t -\frac{v_1}{2 \alpha} \log( \frac{\omegam}{2 \alpha} ) + \frac{\omegam}{2\alpha}, \label{eq:ansatz_for_intercept_exp} \ee 
where the intercept can be understood as $\expval{X}$. Numerics confirm the scaling of intercept in \eqref{eq:ansatz_for_intercept_exp}, although the two-parameter model does not exactly give the numerical value of the intercept.
\eqref{eq:operator_ansatz} implies that $\expval{\hat{n}(t)^2}$ is a quadratic polynomial in $t$. Importantly, \eqref{eq:operator_ansatz} constrains the coefficients of these polynomials in terms of those of $\expval{\hat{n}(t)}^2$. We write a quadratic polynomial ansatz for the variance 
\be \expval{\hat{n}(t)}^2 - \expval{\hat{n}(t)} \approx A_2 t^2 +  A_1 t + A_0.  \label{eq:quadratic_polynomial_model}   \ee 
where $A_2 = v_2-v^2_1$. $A_0$ can be approximated by evaluating \eqref{eq:quadratic_polynomial_model} at $t=t_{\text{on}}$ using the answer from the exponential growth regime, that gives $A_0 \approx e^{4 \alpha t_{\text{on}}} = (\frac{\omegam}{\alpha})^2$. We empirically find that $A_{1}<0$ and $ A_2 \ll \abs{A_1} \ll A_0$, which we can attribute to small \textit{negative} covariance of $B_{n_c}$ with $X$. This hierarchy of coefficients predicts approximate $\sqrt{A_0}/(v_1 t)$ decay for parametrically long times and a small negative drift below  $\sqrt{v_2-v^2_1}/v_1$, before $\delta n(t)$ reaches saturation. These features are confirmed in Fig.~\ref{fig:xy_model} and Fig.~\ref{fig:varying_omegam}. \\~\ 
\sect{The XY Spin Chain}
We consider isotropic XY Hamiltonian in terms of Pauli matrices $\sigma^{\alpha}_j$ at site $j$:
\be H = \frac{1}{4}\sum^{N}_{j=1} ( \sigma^{x}_{j} \sigma^{x}_{j+1}+\sigma^{y}_{j} \sigma^{y}_{j+1}), \ee 
with periodic boundary conditions, and the $N \rightarrow \infty$ limit of the (unnormalized) two-point function
\be C(t) = \frac{1}{4 \tr(e^{-\beta H }  )} \tr(e^{-\beta H/2} \sigma^{z}_{j}(t) e^{-\beta H/2} \sigma^{z}_{j}(0)  ) , \ee 
which was evaluated in \cite{spinhalf, Avdoshkin_2024} as 
\be  C(t) = \frac{1}{4 \pi^2} \bigg( \int^{-1}_{-1} \frac{dx}{\sqrt{1-x^2}} \frac{\cos(t x)}{\cosh(\beta x /2)} \bigg)^2  \label{eq:xy_t_domain}. \ee
Taking the Fourier transform gives the convolution $\Phi(E) = (g_{\beta} *g_{\beta} )(E)$ in terms of
\be g_{\beta}(x) = \frac{\Theta(1-x^2)}{2 \pi \sqrt{1-x^2} \cosh(\beta x/2)}, \ee
where $\Theta$ is the Heaviside function. The claim $\omegam = 2$ follows directly from the support of $g_{\beta}(x)$. \\  
\sect{Numerics using truncated Jacobi matrix}
For numerical computations of complexity, we truncate the matrix $M$ at $n_{\text{cut}}$. Consistency of the answers obtained using the cutoff is enforced by ensuring that $\varphi_{n}(t) \approx 0$ (up to numerical precision) for the sampled $t$, for $n$ much smaller than $n_{\text{cut}}$. \\~\ 

For \eqref{eq:xy_t_domain}, it is straightforward to sample moments with enough accuracy such that $n$ is much larger than $n_c$ and the $b_n$ coefficients have converged to $\omegam/2$ up to small $<10^{-3}$ fluctuations. To access late times to see the saturation of $\delta n(t)$, we padded the numerically computed $M$ with $\omegam/2$ along the off-diagonal. We verified that this padding does not change the time-dependence of $\expval{\hat{n}^k(t)}$ to any observable effect in this regime. 
\clearpage
\setcounter{section}{0}
\setcounter{equation}{0}
\setcounter{figure}{0}
\setcounter{table}{0}
\renewcommand{\thesection}{SM}
\renewcommand{\theequation}{S\arabic{equation}}
\renewcommand{\thefigure}{S\arabic{figure}}
\renewcommand{\thetable}{S\arabic{table}}
\setcounter{secnumdepth}{3}
\begin{center}\textbf{Supplemental Material}\end{center}
\subsection{Krylov amplitudes from saddle point \label{sec:krylov_probabilities}}
The asymptotic form \eqref{eq:szego_asymptotic_pn} provides a good starting point for computing $\varphi_n(t)$ 
\be \varphi_n(t) =  \sqrt{\frac{2 \e0}{\pi}} \int^{\pi}_{0} d\theta \sqrt{\tilde{\Phi}(\theta)} \cos( n \theta +\gamma) e^{-i \e0 t \cos \theta},  \ee 
using the stationary phase method, where we remind that $\tilde{\Phi}(\theta)=\e0 \Phi( \e0 \cos \theta ) \sin(\theta)$. We expand the $\cos( n \theta + \gamma)$ and keep
\be \varphi_n(t) \approx \sqrt{\frac{2 \e0}{\pi}} \int^{\pi}_{0} d\theta \sqrt{\tilde{\Phi}(\theta)}  e^{-i F(\theta) }, \ee
where we reuse $F(\theta)$ to mean
\be F(\theta) = n \theta + \e0 t \cos \theta + \gamma(\theta).  \ee 
Assuming that $\partial_{\theta}\gamma(\theta)$ isn't negative with a very large modulus, the integrand is localized around $\partial_{\theta}F(\theta)=0$ for $t>0$. Now we will further drop $\partial_{\theta}\gamma(\theta)$ as a first approximation. This is not necessarily justified, but is done for the sake of simplicity.  
Then the stationary points are defined by the equation
\be n = \e0 t \sin \theta, \ee
which has solutions $\theta_{1} = \arcsin(n/\e0 t)$ and  $\theta_2 = \pi - \theta_{1}$. Evaluating $\partial^2_{\theta}F(\theta)$ at the stationary points gives $r_1:= \partial^2_{\theta}F(\theta)|_{\theta=\theta_1}=-\sqrt{(\e0 t)^2- (n)^2}$ and $r_2 = -r_1$. The stationary point answer gives 
\be \varphi_n(t) \approx \frac{1}{(\e0^2 t^2 - n^2)^\frac{1}{4}} \bigg[ \sqrt{\tilde{\Phi}(\theta_1)} e^{i \chi } + (-1)^n \sqrt{\tilde{\Phi}(\theta_2)}  e^{-i \chi  } \bigg], \label{eq:two_saddle_stationary} \ee 
in terms of $\chi = - n \theta_1 - \sqrt{\e0^2 t^2 - n^2} + \frac{\pi}{4}$. The overall phase in \eqref{eq:two_saddle_stationary} should not be trusted because $\gamma$ was neglected. \eqref{eq:two_saddle_stationary} diverges as $n = \e0 t$. This signals the breakdown of the Gaussian approximation as the two saddles $\theta_1$ and $\theta_2$ merge at $\frac{\pi}{2}$. Expanding $F(\theta)$ to cubic order around $\theta=\frac{\pi}{2}$ gives the expansion near $n = \e0 t$ 
\be \varphi_n(t) \approx  \frac{e^{- i n \pi/2}}{\sqrt{2 \pi}}   \int \sqrt{\tilde{\Phi}(\theta)} e^{-i (n - \e0 t) (\theta - \frac{\pi}{2} ) + \frac{\e0 t}{6} (\theta - \frac{\pi}{2} )^{3}} d\theta, \label{eq:front_varphi} \ee 
To get a simple analytical approximation for \eqref{eq:front_varphi}, we assume that $\tilde{\Phi}(\theta)$ is slowly varying around $\theta =\frac{\pi}{2}$, so that we can approximate the integral by only keeping $\sqrt{\tilde{\Phi}(\pi/2)}$ in the Taylor expansion. Then the integral over $\theta$ can be cast into the standard Airy integral by extending integration domain to $\mathbb R$. This gives the answer near $n \sim E_0 t$
\be \varphi_n(t) \approx  \sqrt{2 \pi \tilde{\Phi}(\pi/2 )} \bigg( \frac{2}{\e0 t} \bigg)^{\frac{1}{3}}  e^{- i n \frac{\pi}{2}} \text{Ai}\big( -\xi \big) \label{eq:airy_front}, \ee 
where $\xi = \big(\frac{2}{\e0 t} \big)^{\frac{1}{3}}  (\e0 t - n)$.\\
\eqref{eq:airy_front} resembles the continuum limit answers obtained by \cite{Barb_n_2019}. The oscillatory asymptotics of \eqref{eq:airy_front} matches on to \eqref{eq:two_saddle_stationary}. It is easy to see this for the prefactor. To match the phase we can take and $e^{i \chi} + e^{-i \chi} \sim \cos(\chi)$ and then expand in $(\e0 t - n)/\e0 t$ to get $\sin(\frac{2}{3} \xi^{\frac{3}{2}}  + \frac{\pi}{4})$ which matches the oscillatory piece in the expansion of  $\text{Ai}(-\xi)$ for $\xi>0$.  \eqref{eq:airy_front} should be matched on to \eqref{eq:two_saddle_stationary} for some $n_{\text{match}}$ such that $\e0 t - n_{\text{match}} \lesssim O(1) (\e0 t)^{\frac{1}{3}}$. Together \eqref{eq:two_saddle_stationary}, \eqref{eq:airy_front} can provide a remarkably good approximation for $\abs{\varphi_n(t)}^2$.  

\begin{figure}[!htbp]
  \centering
  \includegraphics[width=0.42\textwidth]{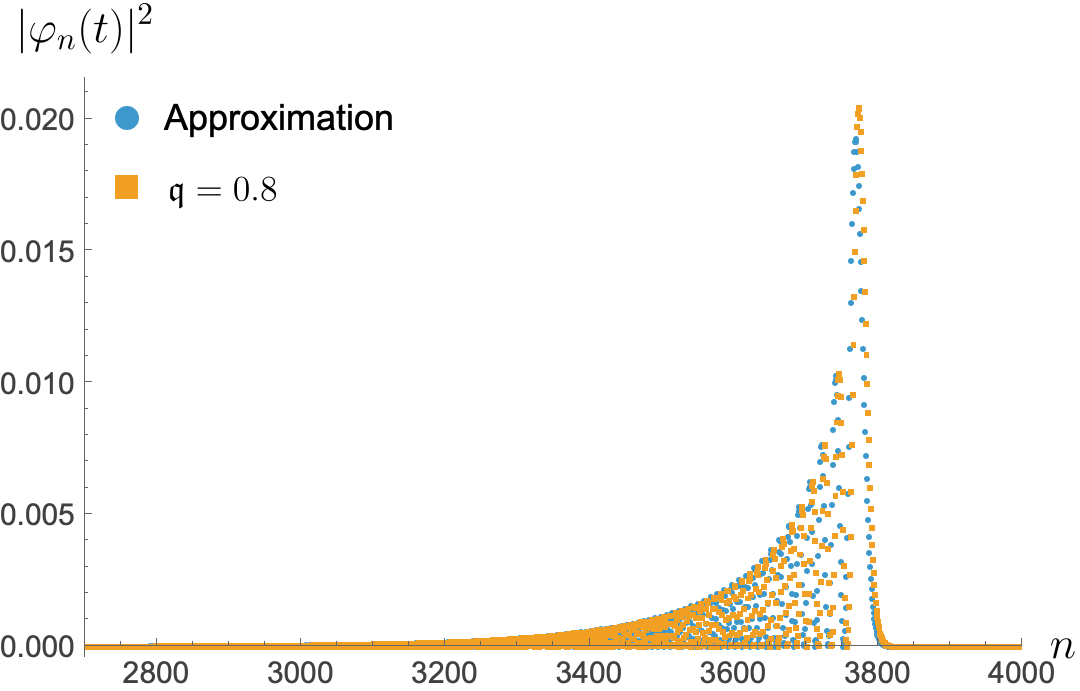}
   \caption{Comparison of $\abs{\varphi_n(t)}^2$ with the combination of \eqref{eq:two_saddle_stationary} and \eqref{eq:airy_front} matched at $n_{\text{match}}=3770$ for DSSYK specified by \eqref{eq:dssyk_lanczos}.  $t=400$ and $\sJ=1$ }   
  \label{fig:asymptotic_phi}
\end{figure}

In Fig.~\ref{fig:asymptotic_phi}, we present an example for DSSYK. A word of caution is that numerics show that the phase $\gamma(\theta)$ can cause important corrections when taking the $\lambda \rightarrow 0$ limit at fixed $t$.  
\subsection{Varying bandwidth}
\begin{figure}[!htbp]
  \centering
  \includegraphics[width=0.38\textwidth]{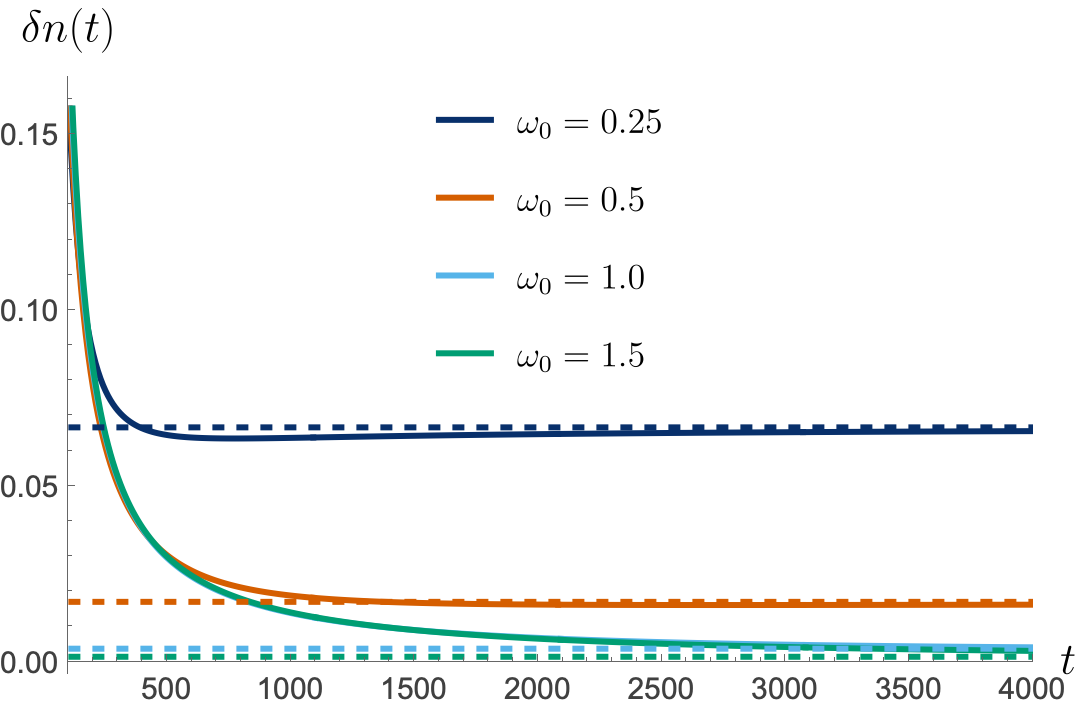}
   \caption{The approach of $\delta n(t)$ for \eqref{eq:fixed_alpha_model} to $\sqrt{v_2 - v^2_1}/v_1 $ with $\alpha = 1/15$. }      
  \label{fig:varying_omegam}
\end{figure}

We numerically studied complexity dynamics for the measure 
\be \Phi(\omega) = \frac{\pi  e^{-\frac{\pi \abs{\omega} }{2 \alpha }}}{4 \alpha 
   \left(1-e^{-\frac{\pi  \omegam}{2 \alpha }}\right)} ~\quad \abs{\omega} \leq \omegam, \label{eq:fixed_alpha_model} \ee 
and $\Phi(\omega) = 0$ for $\omega>\omegam$ to substantiate the model following \eqref{eq:operator_ansatz}. \eqref{eq:fixed_alpha_model} shows a distinct discontinuous edge-behavior. We use it to examine cases where $\alpha$ is held fixed but $\omegam$ is allowed to vary, to complement the study of the XY model spectral function where $\alpha$ varied but $\omegam$ was held fixed. Fig.~\ref{fig:varying_omegam} confirms that $\delta n(t)$ decays as $\frac{1}{\alpha t}$ independent of $\omegam$, before drifting slightly below its infinite time asymptotic. 

\subsection{Finite size signatures \label{sec:finite_size_numerics}}
We numerically study the Hamiltonian
\vspace{-3pt}
\be H=-\sum^{N-1}_{j=1} \sigma^{z}_{j} \sigma^{z}_{j+1} + 1.05 \sum^{j=N}_{j=1} \sigma^{x}_{j} + 0.4 \sum^{j=N-2}_{j=1} \sigma^{x}_{j} \sigma^{x}_{j+1} \sigma^{x}_{j+2}. \label{eq:spin_chain_ham} \ee 
The three-body interaction is chosen to preserve a spectral reflection symmetry while maintaining nonintegrability. For the initial states considered, this symmetry enforces $a_n=0$ reducing overall fluctuations. For the system sizes studied, $11\leq N\leq14$, the expected linear-growth regime is not yet clearly resolved.

\begin{figure}[!htbp]
  \centering
  \includegraphics[width=0.38\textwidth]{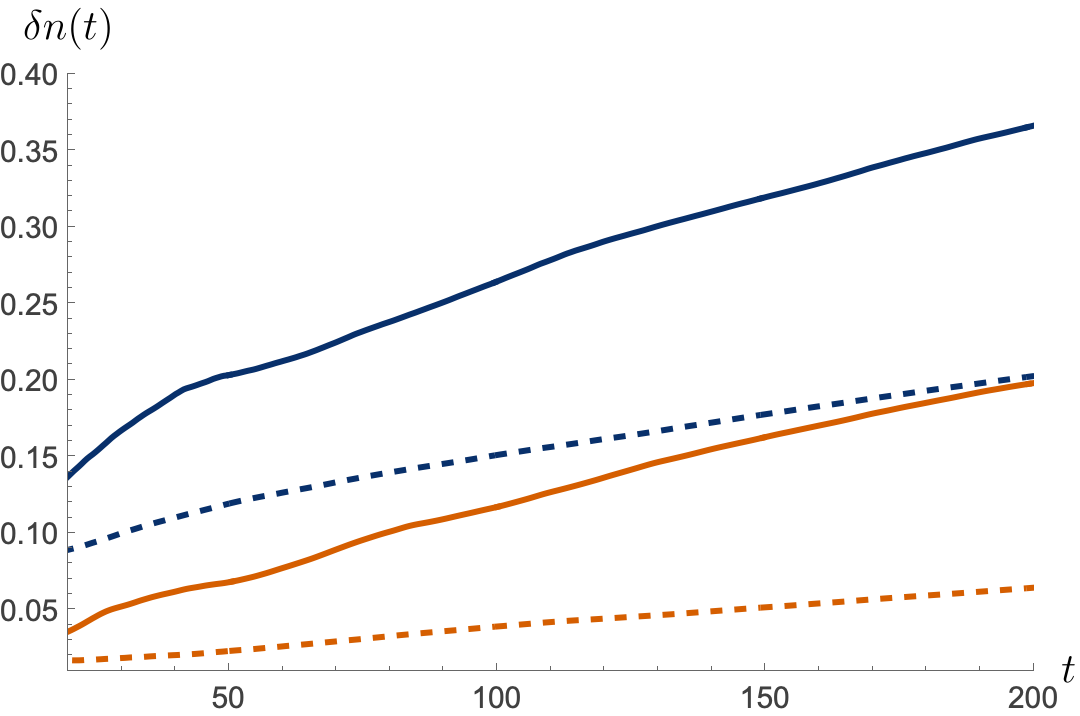}
    \caption{$\delta n(t)$ for $\ket{\psi_1}$ (dark blue) and $\ket{\psi_2}$ (dark orange) evolved by \eqref{eq:spin_chain_ham} for $N=12$ (solid lines) and $N=14$ (dashed lines).}
    \label{fig:spin_chain}
\end{figure}
Nonetheless $\delta n(t)$ shows a clear trend that we illustrate with two examples 
\be \ket{\psi_1} = \frac{1}{2^{N/2}} \sum^{2^{N}}_{n=1} \ket{E_n},~\quad  \ket{\psi_2} \propto \sum^{2^{N}}_{n=1}  e^{-E^2_{n}/(32)} \ket{E_n}. \ee 
The density of states $\rho(E)$ is approximately Gaussian towards the center of the spectrum with variance growing with $N$. For $\ket{\psi_1}$, $\Phi(E)=\rho(E)$, while for $\ket{\psi_2}$, $\Phi(E)$ has a fixed width $\sigma=4$ independent of $N$. Since the spectral bandwidth grows linearly with $N$, the width of $\Phi(E)$ relative to the bandwidth scales as $N^{-1/2}$ and $N^{-1}$, respectively.  

Fig.~\ref{fig:spin_chain} shows $\delta n(t)$ for a time window that could become the linear growth regime in larger systems.  $\delta n(t)$ changes slowly over this window and is seen to become flatter for larger $N$. $\delta n$ is inversely correlated with the width of $\Phi(E)$ rescaled by the bandwidth. This correlation persists across different Hamiltonians and initial states, but eventually saturates when the width $\sigma$ becomes sufficiently small. 
 \end{document}